\documentclass[10pt,a4paper,onecolumn]{article}
\usepackage{marginnote}
\usepackage{graphicx}
\usepackage{xcolor}
\usepackage{authblk,etoolbox}
\usepackage{titlesec}
\usepackage{calc}
\usepackage{tikz}
\usepackage{hyperref}
\hypersetup{colorlinks,breaklinks=true,
            urlcolor=[rgb]{0.0, 0.5, 1.0},
            linkcolor=[rgb]{0.0, 0.5, 1.0}}
\usepackage{caption}
\usepackage{tcolorbox}
\usepackage{amssymb,amsmath}
\usepackage{ifxetex,ifluatex}
\usepackage{seqsplit}
\usepackage{xstring}

\usepackage{float}
\let\origfigure\figure
\let\endorigfigure\endfigure
\renewenvironment{figure}[1][2] {
    \expandafter\origfigure\expandafter[H]
} {
    \endorigfigure
}

\usepackage{fixltx2e} 
\usepackage[
  backend=biber,
]{biblatex}
\bibliography{paper.bib}

\let\textttOrig=\texttt
\def\texttt#1{\expandafter\textttOrig{\seqsplit{#1}}}
\renewcommand{\seqinsert}{\ifmmode
  \allowbreak
  \else\penalty6000\hspace{0pt plus 0.02em}\fi}

\makeatletter
\let\href@Orig=\href
\def\href@Urllike#1#2{\href@Orig{#1}{\begingroup
    \def\Url@String{#2}\Url@FormatString
    \endgroup}}
\def\href@Notdoi#1#2{\def\tempa{#1}\def\tempb{#2}%
  \ifx\tempa\tempb\relax\href@Urllike{#1}{#2}\else
  \href@Orig{#1}{#2}\fi}
\def\href#1#2{%
  \IfBeginWith{#1}{https://doi.org}%
  {\href@Urllike{#1}{#2}}{\href@Notdoi{#1}{#2}}}
\makeatother

\newlength{\cslhangindent}
\newlength{\csllabelwidth}
\newenvironment{CSLReferences}[3] 
 {
  \setlength{\parindent}{0pt}
  \ifodd #1 \everypar{\setlength{\hangindent}{\cslhangindent}}\ignorespaces\fi
  \ifnum #2 > 0
  \setlength{\parskip}{#2\baselineskip}
  \fi
 }%
 {}
\usepackage{calc}

\usepackage[top=3.5cm, bottom=3cm, right=1.5cm, left=1.0cm,
            headheight=2.2cm, reversemp, includemp, marginparwidth=4.5cm]{geometry}

\titleformat{\section}
  {\normalfont\sffamily\Large\bfseries}
  {}{0pt}{}
\titleformat{\subsection}
  {\normalfont\sffamily\large\bfseries}
  {}{0pt}{}
\titleformat{\subsubsection}
  {\normalfont\sffamily\bfseries}
  {}{0pt}{}
\titleformat*{\paragraph}
  {\sffamily\normalsize}

\usepackage{fancyhdr}
\makeatletter
\let\ps@plain\ps@fancy
\fancyheadoffset[L]{4.5cm}
\fancyfootoffset[L]{4.5cm}

\definecolor{linky}{rgb}{0.0, 0.5, 1.0}

\newtcolorbox{repobox}
   {colback=red, colframe=red!75!black,
     boxrule=0.5pt, arc=2pt, left=6pt, right=6pt, top=3pt, bottom=3pt}

\patchcmd{\@maketitle}{center}{flushleft}{}{}
\patchcmd{\@maketitle}{center}{flushleft}{}{}
\patchcmd{\@maketitle}{\LARGE}{\LARGE\sffamily}{}{}
\def\maketitle{{%
  
  \AB@maketitle}}
\makeatletter
\renewcommand\AB@affilsepx{ \protect\Affilfont}
\renewcommand\AB@affilnote[1]{{\bfseries #1}\hspace{3pt}}
\renewcommand{\affil}[2][]%
   {\newaffiltrue\let\AB@blk@and\AB@pand
      \if\relax#1\relax\def\AB@note{\AB@thenote}\else\def\AB@note{#1}%
        \setcounter{Maxaffil}{0}\fi
        \begingroup
        \let\href=\href@Orig
        \let\texttt=\textttOrig
        \let\protect\@unexpandable@protect
        \def\thanks{\protect\thanks}\def\footnote{\protect\footnote}%
        \@temptokena=\expandafter{\AB@authors}%
        {\def\\{\protect\\\protect\Affilfont}\xdef\AB@temp{#2}}%
         \xdef\AB@authors{\the\@temptokena\AB@las\AB@au@str
         \protect\\[\affilsep]\protect\Affilfont\AB@temp}%
         \gdef\AB@las{}\gdef\AB@au@str{}%
        {\def\\{, \ignorespaces}\xdef\AB@temp{#2}}%
        \@temptokena=\expandafter{\AB@affillist}%
        \xdef\AB@affillist{\the\@temptokena \AB@affilsep
          \AB@affilnote{\AB@note}\protect\Affilfont\AB@temp}%
      \endgroup
       \let\AB@affilsep\AB@affilsepx
}
\makeatother

\renewcommand\Affilfont{\sffamily\small\mdseries}
\ifnum 0\ifxetex 1\fi\ifluatex 1\fi=0 
  \usepackage[T1]{fontenc}
  \usepackage[utf8]{inputenc}

\else 
  \ifxetex
    \usepackage{mathspec}
    \usepackage{fontspec}

  \else
    \usepackage{fontspec}
  \fi
  \defaultfontfeatures{Ligatures=TeX,Scale=MatchLowercase}

\fi
\IfFileExists{upquote.sty}{\usepackage{upquote}}{}
\IfFileExists{microtype.sty}{%
\usepackage{microtype}
\UseMicrotypeSet[protrusion]{basicmath} 
}{}

\usepackage{hyperref}
\hypersetup{unicode=true,
            pdftitle={CWS-PResUNet: Music Source Separation with Channel-wise Subband Phase-aware ResUNet},
            pdfborder={0 0 0},
            breaklinks=true}
\let\addcontentslineOrig=\addcontentsline
\def\addcontentsline#1#2#3{\bgroup
  \let\texttt=\textttOrig\addcontentslineOrig{#1}{#2}{#3}\egroup}
\let\markbothOrig\markboth
\def\markboth#1#2{\bgroup
  \let\texttt=\textttOrig\markbothOrig{#1}{#2}\egroup}
\let\markrightOrig\markright
\def\markright#1{\bgroup
  \let\texttt=\textttOrig\markrightOrig{#1}\egroup}

\usepackage{graphicx,grffile}
\makeatletter
\def\maxwidth{\ifdim\Gin@nat@width>\linewidth\linewidth\else\Gin@nat@width\fi}
\def\maxheight{\ifdim\Gin@nat@height>\textheight\textheight\else\Gin@nat@height\fi}
\makeatother
\setkeys{Gin}{width=\maxwidth,height=\maxheight,keepaspectratio}
\IfFileExists{parskip.sty}{%
\usepackage{parskip}
}{
\setlength{\parindent}{0pt}
\setlength{\parskip}{6pt plus 2pt minus 1pt}
}
\ifx\paragraph\undefined\else
\let\oldparagraph\paragraph
\renewcommand{\paragraph}[1]{\oldparagraph{#1}\mbox{}}
\fi
\ifx\subparagraph\undefined\else
\let\oldsubparagraph\subparagraph
\renewcommand{\subparagraph}[1]{\oldsubparagraph{#1}\mbox{}}
\fi

\title{CWS-PResUNet: Music Source Separation with Channel-wise Subband
Phase-aware ResUNet}

        \author[1, 2]{Haohe Liu}
          \author[1]{Qiuqiang Kong}
          \author[1]{Jiafeng Liu}
    
      \affil[1]{Sound, Audio, and Music Intelligence (SAMI) Group,
ByteDance}
      \affil[2]{The Ohio State University}
  \date{\vspace{-7ex}}

\begin{document}
\maketitle

\marginpar{

  \begin{flushleft}
  \sffamily\small

  \vspace{2mm}

  \par\noindent\hrulefill\par

  \vspace{2mm}

  \vspace{2mm}
  {\bfseries License}\\
  Authors of papers retain copyright and release the work under a Creative Commons Attribution 4.0 International License (\href{http://creativecommons.org/licenses/by/4.0/}{\color{linky}{CC BY 4.0}}).

  \vspace{4mm}
  {\bfseries In partnership with}\\
  \vspace{2mm}
  \includegraphics[width=4cm]{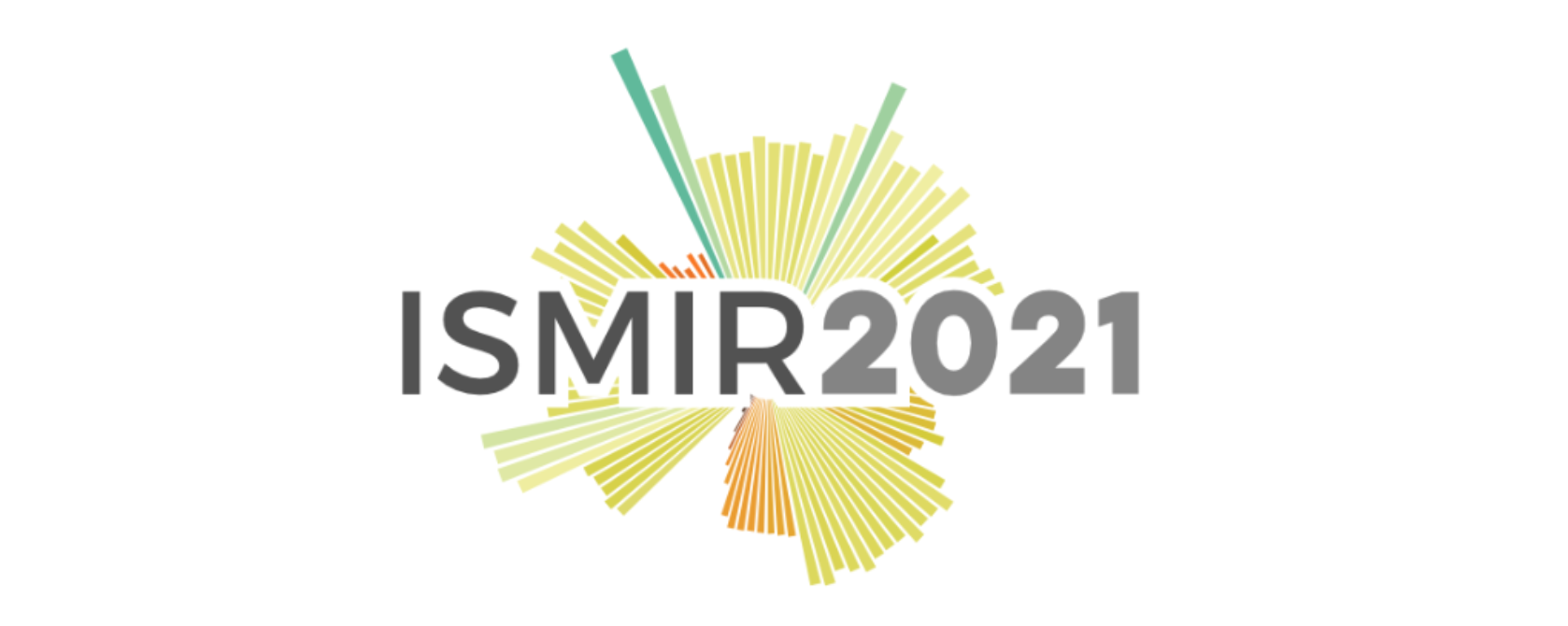}

  \end{flushleft}
}

\hypertarget{abstract}{%
\section{Abstract}\label{abstract}}

Music source separation (MSS) shows active progress with deep learning
models in recent years. Many MSS models perform separations on
spectrograms by estimating bounded ratio masks and reusing the phases of
the mixture. When using convolutional neural networks (CNN), weights are
usually shared within a spectrogram during convolution regardless of the
different patterns between frequency bands. In this study, we propose a
new MSS model, channel-wise subband phase-aware ResUNet (CWS-PResUNet),
to decompose signals into subbands and estimate an unbound complex ideal
ratio mask (cIRM) for each source. CWS-PResUNet utilizes a channel-wise
subband (CWS) feature to limit unnecessary global weights sharing on the
spectrogram and reduce computational resource consumptions. The saved
computational cost and memory can in turn allow for a larger
architecture. On the MUSDB18HQ test set, we propose a 276-layer
CWS-PResUNet and achieve state-of-the-art (SoTA) performance on
\texttt{vocals} with an 8.92 signal-to-distortion ratio (SDR) score. By
combining CWS-PResUNet and Demucs, our ByteMSS system ranks the 2nd on
\texttt{vocals} score and 5th on average score in the 2021 ISMIR Music
Demixing (MDX) Challenge limited training data track (leaderboard A).
Our code and pre-trained models are publicly available\footnote{Open
  sourced at:
  https://github.com/haoheliu/2021-ISMIR-MSS-Challenge-CWS-PResUNet}.

\hypertarget{introductions-and-related-works}{%
\section{Introductions and Related
Works}\label{introductions-and-related-works}}

Music source separation aims at decomposing a music mixture into several
soundtracks, such as \texttt{Vocals}, \texttt{Bass}, \texttt{Drums}, and
\texttt{Other} tracks. It is closely related to topics like music
transcription, remixing, and retrieval. Based on deep learning models,
most of the early studies (Jansson et al., 2017; Takahashi et al., 2018)
perform separations in the frequency domain by estimating the ideal
ratio masks (IRM) of the magnitude spectrogram and reusing the phase of
the mixture. Later, time-domain models (Défossez et al., 2019) start to
demonstrate SoTA performance using direct waveform modeling, which does
not involve transformations like short-time fourier transform (STFT). In
this case, phase information can be implicitly estimated and models will
not be restricted with the fixed time-frequency resolution. To enhance
the MSS performance, Y. Liu et al. (2020) chose to employ a
self-attension mechanism and Dense-UNet architecture. Choi et al. (2019)
compared the performance of several types of UNet built with different
intermediate blocks. To alleviate the computational cost, Kadandale et
al. (2020) designed a multi-task model to replace source-dedicated
models. Also, H. Liu et al. (2020) proposed to use the channel-wise
subband feature to reduce resource consumptions and improve separation
performance. Recently, Kong et al. (2021) conducted an experiment on the
MSS system theoretical upper bound, which proved the limitation of IRMs
and the importance of phase estimation.

In the next section, we will introduce the detailed architecture of
CWS-PResUNet as well as ByteMSS, the system we submitted for the MDX
Challenge (Mitsufuji et al., 2021).

\begin{figure}
\centering
\includegraphics[width=1\textwidth,height=\textheight]{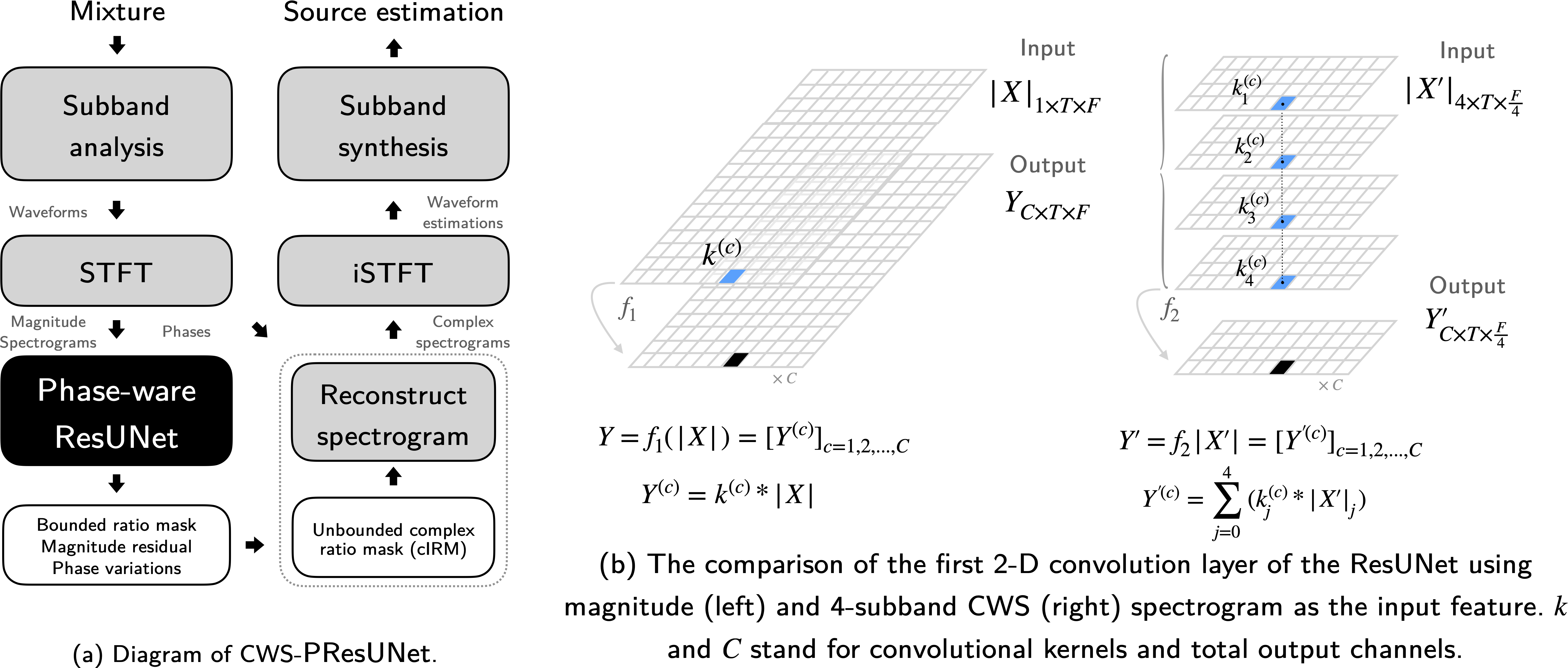}
\caption[Overview of the CWS-PResUNet and a comparison between using
magnitude spectrogram and channel-wise subband spectrogram as the input
feature.]{Overview of the CWS-PResUNet and a comparison between using
magnitude spectrogram and channel-wise subband spectrogram as the input
feature.\footnotemark{}}
\end{figure}
\footnotetext{We use mono signal for simple illustration.}

\hypertarget{method}{%
\section{Method}\label{method}}

CWS-PResUNet is a ResUNet (H. Liu et al., 2021) based model integrating
the CWS feature (H. Liu et al., 2020) and the cIRM estimation strategies
described in Kong et al. (2021). The overall pipeline is summarized in
Figure 1a. We modeling separation on the subband spectrogram and phase
domains. The analysis and synthesis filters in subband operations are
designed by optimizing reconstruction error using the open-source
toolbox\footnote{https://www.mathworks.com/matlabcentral/fileexchange/40128-filter-bank-design}.

As is illustrated in Figure 1b, the CWS feature has a lower frequency
dimension and more channels compared with the full band spectrogram. To
adapt the conventional full band CNN-based model to the CWS input
feature, it just needs to modify the input and final output channel with
internal CNN blocks unchanged. In this way, the internal feature map of
the model becomes smaller, leading to a direct reduction in
computational cost. Also, models become more efficient by enlarging
receptive fields and diverging subband information into different
channels.

The detailed computation procedure of our CWS-PResUNet model is
described as follows. For a stereo mixture signal
\(x \in \mathbb{R}^{2\times L}\), where \(L\) stands for signal length,
we first utilize a set of analysis filters \({h}^{(j)},j=1,2,3,4\) to
perform subband decompositions: \[
x^{\prime}_{8\times \frac{L}{4}} = [\text{DS}_4({x_{2\times 1 \times L}}*{h}^{(j)}_{1\times 64})]_{j=1,2,3,4},
\] where \(\text{DS}_{4}(\cdot)\), \(*\), and \([\cdot]\) denote the
downsampling by 4, convolution, and stacking operators, respectively.
The analysis filters we used are uniform filter banks with a filter
length of 64. Then we calculate the STFT of the downsampled subband
signals \(x^{\prime}\) to obtain their magnitude spectrograms
\(|X^{\prime}|_{8\times T \times \frac{F}{4}}\), which is the input of
Phase-aware ResUNet.

\begin{figure}
\centering
\includegraphics[width=1\textwidth,height=\textheight]{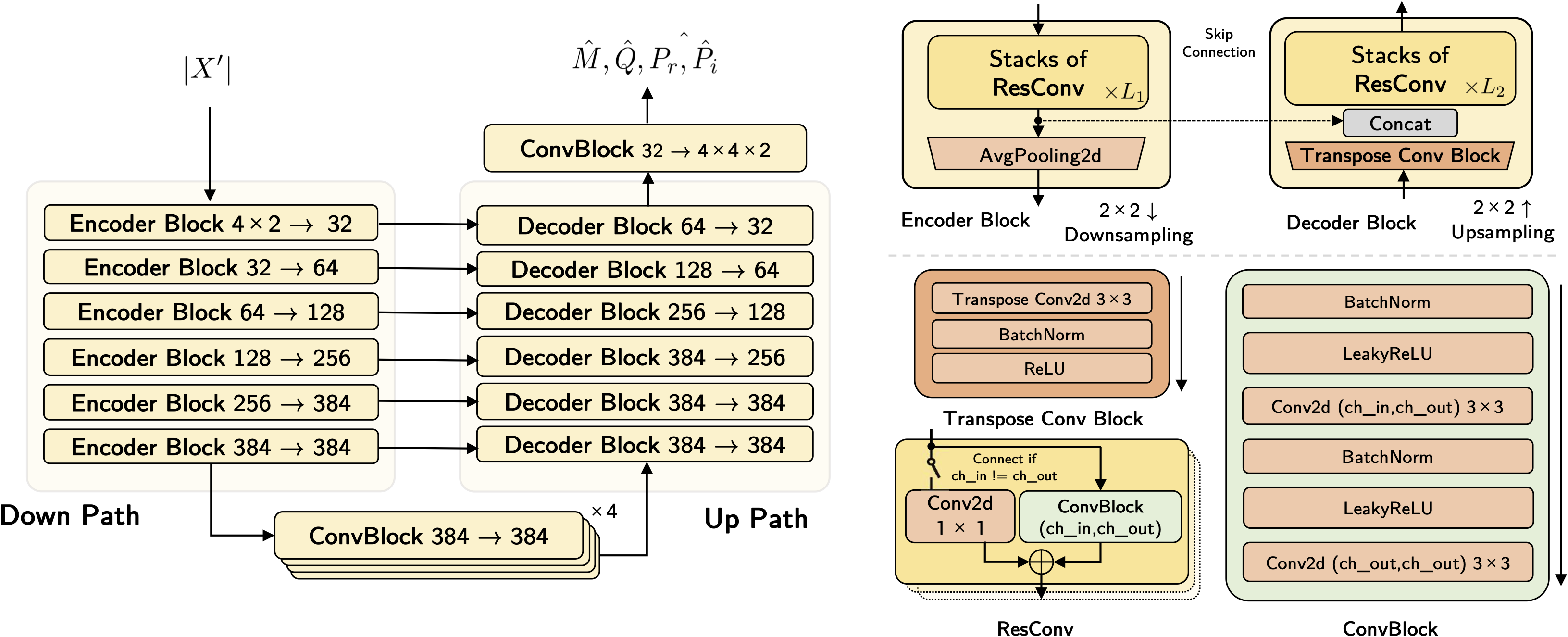}
\caption{The architecture of Phase-aware ResUNet}
\end{figure}

As is shown in Figure 2, the phase-aware ResUNet is a symmetric
architecture containing a down-sampling and an up-sampling path with
skip-connections between the same level. It accepts \(|X^{\prime}|\) as
input and estimates four tensors with the same shape: mask estimation
\(\hat{M}\), phase variation \(\hat{P}_{r}\), \(\hat{P}_{i}\), and
direct magnitude prediction \(\hat{Q}\). The complex spectrogram can be
reconstructed with the following equation: \[
\hat{S}^{\prime} = \text{relu}(|X^{\prime}|\odot \text{sigmoid}(\hat{M})+\hat{Q})\exp^{j(\angle X^{\prime} +\angle \hat{\theta})},
\] in which
\(cos\angle \hat{\theta}=\hat{P}_{r}/(\sqrt{\hat{P}_{r}^2+\hat{P}_{i}^2})\)
and
\(sin\angle \hat{\theta}=\hat{P}_{i}/(\sqrt{\hat{P}_{r}^2+\hat{P}_{i}^2})\).
We pass the mask estimation \(\hat{M}\) through a sigmoid function to
obtain a mask with values between 0 and 1. Then by estimating
\(\hat{Q}\) and \(\hat{\theta}\), models can avoid using mixture phase
and estimating mask with only bounded values to calculate the unbounded
cIRM. We use relu activation to ensure the positve magnitude value.
Finally, after the inverse STFT, we perform subband reconstructions to
obtain the source estimation \(\hat{s}\): \[ 
\hat{s}_{2\times L} = \sum_{j=1}^{4}(\text{US}_4(\hat{s}^{\prime}_{2\times 4\times \frac{L}{4}})*g^{(j)}_{4\times 64}),
\] where \(g^{(j)}, j=1,2,3,4\) are the pre-defined synthesis filters
and \(\text{US}_4(\cdot)\) is the zero-insertion upsampling function.

Our model for \texttt{vocals} is optimized by calculating L1 loss
between \(\hat{s}\) and its target source \(s\). Although we also use a
model dedicated to separating the \texttt{other} track, we notice
estimating and optimizing four sources together in one model can result
in a 0.2 SDR (Vincent et al., 2006) gain on \texttt{other}. In this
case, we not only use L1 loss on the waveform, but also employ
energy-conservation loss, which calculates the L1 loss between the
mixture and the sum of four source estimations. Our CWS-PResUNet models
for \texttt{bass} and \texttt{drums} reported in the next section employ
the same setup as the model for \texttt{other}.

In our ByteMSS system, we set up Demucs (Défossez et al., 2019) to
separate \texttt{bass} and \texttt{drums} tracks because it performs
better than CWS-PResUNet on these two sources. Demucs is a time-domain
MSS model. In our study, we adopted the open-sourced pre-trained
Demucs\footnote{https://github.com/facebookresearch/demucs} and do not
apply the shift trick because it will slow down the inference speed. To
separate the \texttt{vocals} track, we train a 276-layer CWS-PResUNet.
For the \texttt{other} track, which is usually prone to overfitting due
to the limited training data, we setup a smaller 166-layer CWS-PResUNet
in order to achieve better generalization ability on the hidden test
set.

\hypertarget{experiments}{%
\section{Experiments}\label{experiments}}

Our models are optimized using the training subset of MUSDB18HQ (Rafii
et al., 2019). We calculate the STFT of the downsampled 11.05 kHz
subband signals with a window length of 512 and a window shift of 110.
We use Adam optimizer with an initial learning rate of 0.001 and
exponential decay. CWS-PResUNet takes approximately four days to train
on a Tesla V100 GPU. During inference, we utilize a 10-second long
\texttt{boxcar} windowing function (Schuster et al., 2008) with no
overlapping to segment the signal. For evaluation, we report the SDR on
the MUSDB18HQ test set with the open-sourced \emph{museval} tool (Stöter
et al., 2018).

The subband analysis and synthesis operations usually cannot achieve
perfect reconstruction. To assess the errors introduced by subband
operations, we decompose the test set \texttt{vocals} tracks into 2,4,
and 8 subbands and reconstruct them back to evaluate the reconstruction
error of the filterbanks. We perform the computation using 32 bits float
numbers. As is presented in Table 1, in all cases subband
reconstructions achieve high performance with neglectable errors, which
show an increasing trend with more subband numbers.

\includegraphics[width=1\textwidth,height=\textheight]{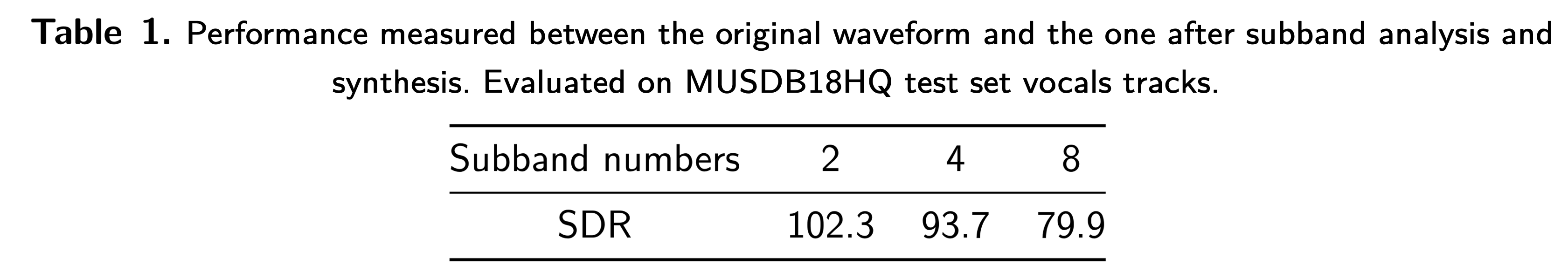}

Table 2 lists the results of the baselines and our proposed systems. Our
CWS-PResUNets achieve an SDR of 8.92 and 5.84 on \texttt{vocals} and
\texttt{other} sources, respectively, outperforming the baseline X-UMX
(Sawata et al., 2021), D3Net (Takahashi \& Mitsufuji, 2020), and Demucs
systems by a large margin. Demucs performs better than CWS-PResUNet on
\texttt{bass} and \texttt{drums} tracks. We assume that is because
time-domain models can learn better representations than time-frequency
features so are more suitable for separating percussive and band-limited
sources. The average performance of our ByteMSS system is 6.97, marking
a SoTA performance on MSS. Considering the high performance of the
\texttt{vocals} model, we also attempt to separate three instrumental
sources from \texttt{mixture} minus \texttt{vocals}. In this case, the
average score remains 6.97, in which the \texttt{drums} score increase
to 6.72 but the other three sources drop slightly. In the future, we
will address the integration of time and frequency models for the
compensations in both domains.

\includegraphics[width=1\textwidth,height=\textheight]{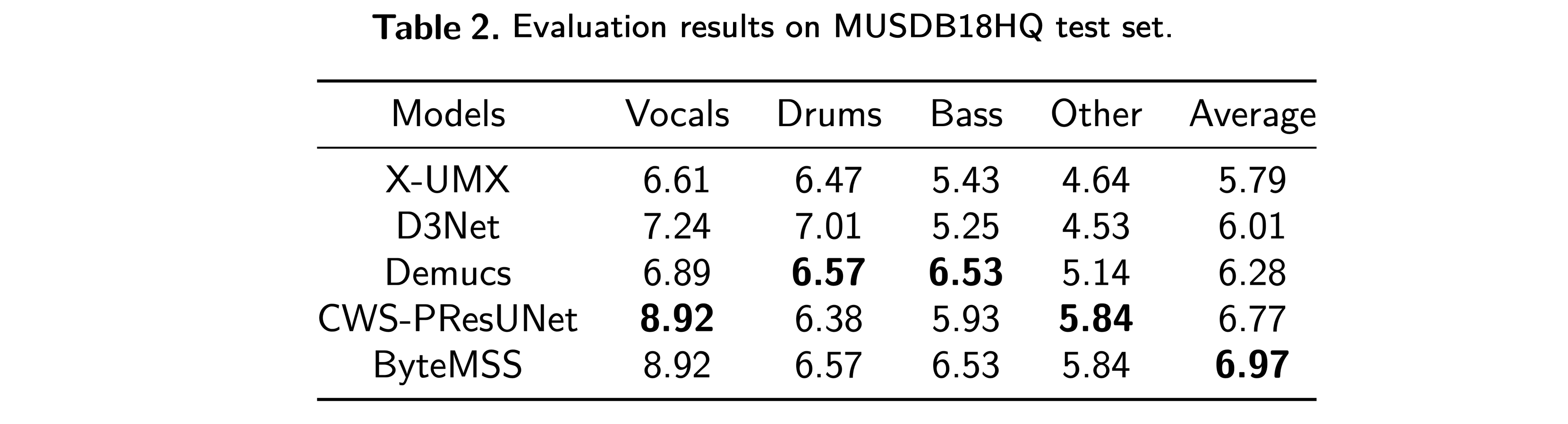}

\hypertarget{conclusions}{%
\section{Conclusions}\label{conclusions}}

Our experiment result shows CWS-PResUNet can achieve a leading
performance on the separation of \texttt{vocals} and \texttt{other}
tracks. And channel-wise subband feature is an effective alternative to
magnitude spectrogram on music source separation task.

\hypertarget{acknowledgements}{%
\section{Acknowledgements}\label{acknowledgements}}

This project is funded by ByteDance Inc.~We acknowledge the supports
from Haonan Chen for testing our system.

\hypertarget{reference}{%
\section*{Reference}\label{reference}}
\addcontentsline{toc}{section}{Reference}

\hypertarget{refs}{}
\begin{CSLReferences}{1}{0}
\leavevmode\hypertarget{ref-choi2019investigating}{}%
Choi, W., Kim, M., Chung, J., Lee, D., \& Jung, S. (2019). Investigating
u-nets with various intermediate blocks for spectrogram-based singing
voice separation. \emph{arXiv Preprint arXiv:1912.02591}.

\leavevmode\hypertarget{ref-defossez2019demucs}{}%
Défossez, A., Usunier, N., Bottou, L., \& Bach, F. (2019). Demucs: Deep
extractor for music sources with extra unlabeled data remixed.
\emph{arXiv Preprint arXiv:1909.01174}.

\leavevmode\hypertarget{ref-jansson2017singing}{}%
Jansson, A., Humphrey, E., Montecchio, N., Bittner, R., Kumar, A., \&
Weyde, T. (2017). \emph{Singing voice separation with deep u-net
convolutional networks}.

\leavevmode\hypertarget{ref-kadandale2020multi}{}%
Kadandale, V. S., Montesinos, J. F., Haro, G., \& Gómez, E. (2020).
\emph{Multi-task u-net for music source separation}.

\leavevmode\hypertarget{ref-kong2021decoupling}{}%
Kong, Q., Cao, Y., Liu, H., Choi, K., \& Wang, Y. (2021). Decoupling
magnitude and phase estimation with deep ResUNet for music source
separation. \emph{arXiv Preprint arXiv:2109.05418}.

\leavevmode\hypertarget{ref-liu2021voicefixer}{}%
Liu, H., Kong, Q., Tian, Q., Zhao, Y., Wang, D., Huang, C., \& Wang, Y.
(2021). VoiceFixer: Toward general speech restoration with neural
vocoder. \emph{arXiv Preprint arXiv:2109.13731}.

\leavevmode\hypertarget{ref-liu2020channel}{}%
Liu, H., Xie, L., Wu, J., \& Yang, G. (2020). Channel-wise subband input
for better voice and accompaniment separation on high resolution music.
\emph{arXiv Preprint arXiv:2008.05216}.

\leavevmode\hypertarget{ref-liu2020voice}{}%
Liu, Y., Thoshkahna, B., Milani, A., \& Kristjansson, T. (2020). Voice
and accompaniment separation in music using self-attention convolutional
neural network. \emph{arXiv Preprint arXiv:2003.08954}.

\leavevmode\hypertarget{ref-mitsufuji2021music}{}%
Mitsufuji, Y., Fabbro, G., Uhlich, S., \& Stöter, F.-R. (2021). Music
demixing challenge at ISMIR 2021. \emph{arXiv Preprint
arXiv:2108.13559}.

\leavevmode\hypertarget{ref-rafii2019musdb18}{}%
Rafii, Z., Liutkus, A., Stöter, F.-R., Mimilakis, S. I., \& Bittner, R.
(2019). \emph{MUSDB18-HQ-an uncompressed version of MUSDB18}.

\leavevmode\hypertarget{ref-x-umx-sawata2021all}{}%
Sawata, R., Uhlich, S., Takahashi, S., \& Mitsufuji, Y. (2021). All for
one and one for all: Improving music separation by bridging networks.
\emph{IEEE International Conference on Acoustics, Speech and Signal
Processing}, 51--55.

\leavevmode\hypertarget{ref-schuster2008influence}{}%
Schuster, S., Scheiblhofer, S., \& Stelzer, A. (2008). The influence of
windowing on bias and variance of DFT-based frequency and phase
estimation. \emph{IEEE Transactions on Instrumentation and Measurement},
1975--1990.

\leavevmode\hypertarget{ref-SiSEC18}{}%
Stöter, F.-R., Liutkus, A., \& Ito, N. (2018). \emph{The signal
separation evaluation campaign}. 293--305.

\leavevmode\hypertarget{ref-takahashi2018mmdenselstm}{}%
Takahashi, N., Goswami, N., \& Mitsufuji, Y. (2018). {MMDenseLSTM}: An
efficient combination of convolutional and recurrent neural networks for
audio source separation. \emph{International Workshop on Acoustic Signal
Enhancement}, 106--110.

\leavevmode\hypertarget{ref-takahashi2020d3net}{}%
Takahashi, N., \& Mitsufuji, Y. (2020). D3net: Densely connected
multidilated densenet for music source separation. \emph{arXiv Preprint
arXiv:2010.01733}.

\leavevmode\hypertarget{ref-vincent2006performance}{}%
Vincent, E., Gribonval, R., \& Févotte, C. (2006). Performance
measurement in blind audio source separation. \emph{IEEE Transactions on
Audio, Speech, and Language Processing}, 1462--1469.

\end{CSLReferences}

\end{document}